\documentclass{sig-alternate}
\usepackage{amsmath,amssymb,amsfonts} 
\usepackage[T1]{fontenc} 
\usepackage{url}
\usepackage{graphicx}
\usepackage{epsfig}
\usepackage{multirow}
\usepackage[boxed,ruled,noend,vlined,linesnumbered]{algorithm2e}
\usepackage[noend]{algorithmic}
\usepackage{subfigure}
\usepackage{stmaryrd}
\usepackage{lipsum}
\newcommand{\ur}{\texttt{u}}

\newcommand{\q}{\texttt{q}}
\renewcommand{\a}{\texttt{a}}

\newcommand{\e}{\texttt{e}}
\newcommand{\dbpo}{\texttt{dbpedia.org}}
\newcommand{\fbo}{\texttt{freebase.org}}
\newcommand{\geoo}{\texttt{geonames.org}}
\newcommand{\yagoo}{\texttt{yago.org}}
\newcommand{\nautilod}{\textsc{NautiLOD}}
\newcommand{\swget}{\texttt{swget}}
\newcommand{\dswget}{\texttt{Dswget}}

\newcommand{\UU}{{\mathcal U}}

\newcommand{\LL}{{\mathcal L}}

\newcommand{\DD}{{\mathcal D}}
\newcommand{\TT}{{\mathcal T}}

\renewcommand{\DD}{{\mathcal D}}

\newcommand{\WW}{\mathcal{W}}

\newcommand{\tuple}[1]{\langle #1 \rangle}

\newcommand{\pathw}{\texttt{path}}

\newtheorem{theorem}{Theorem}[section]

\newtheorem{defn}[theorem]{\textbf{Definition}}
\newtheorem{exmp}[theorem]{\textbf{Example}}

\begin{document}
%%%
%%
\title{Semantic Navigation on the Web of Data:
Specification of Routes, Web Fragments and Actions}
\numberofauthors{3}
\author{
\alignauthor
Valeria Fionda\\
       \affaddr
       {KRDB, Free University of Bozen-Bolzano, Italy}
 %      \email   {fionda@inf.unibz.it}
       \alignauthor
       Claudio  Gutierrez\\
       \affaddr
       {DCC, Universidad de Chile, Santiago, Chile}
%       \email       {cgutierr@dcc.uchile.cl}
\alignauthor 
Giuseppe Pirr\'o\\
       \affaddr
       {KRDB, Free University of Bozen-Bolzano, Italy}
%       \email
%       {pirro@inf.unibz.it}
}
\maketitle
%%%%% ABSTRACT%%%%%%%%%%%%%%%%%%%%%%
\begin{abstract}
 The massive semantic data sources linked in the Web of Data
 give new meaning to old features like navigation; 
introduce new challenges like semantic specification 
of Web fragments; and make it possible to specify
actions relying on semantic data. In this paper we introduce a declarative
 language to face these challenges. Based on navigational
features, it is designed to specify fragments of the Web of Data and
actions to be performed based on these data.
  We implement it in a centralized fashion, 
and show its power and performance. Finally, we
 explore the same ideas in a distributed setting, showing their
 feasibility, potentialities and challenges.
\vspace{-.1cm}
\end{abstract}
\keywords{\vspace{-.1cm} Navigation, Web of Data, Linked Data, Semantic Web}
\section{Introduction} 
%%
%%%%%%%%%%%%%%%%%%% The setting
%%
Classically the Web has been modelled as a huge graph of links between pages~\cite{Brin1998}. This model included Web features such as links without labels and {\em only} generated by the owner of the page.\footnote{Even though the spec. XLink~\cite{DeRose2010} allows to define links in a third page, it was never used massively.} 
Although  Web pages are created and kept distributively, 
their small size and lack of structure stimulated the idea to view searching and querying through single and centralized repositories 
(built from pages via crawlers). With the advent of the Web of Data, that is, semantic data at massive scale~\cite{bizer09a,Heath2011}, these assumptions, in general, do not hold anymore. First, links are semantically labelled (thanks to RDF triples) thus can be used to orient and control the navigation, are generated distributively and can be part of any data source. Hence, it has become a reality --using the words of Tim Berners-Lee-- that {\em anyone} can say {\em anything} about {\em anything} and publish it {\em anywhere}. Second, data sources have a truly distributed nature due to their huge size, autonomous generation, and standard RDF structure. This makes inconvenient and impractical to re-organize them in central repositories as for Web pages.

\begin{figure}[!t]
\centering
 \vspace{-.3cm}
 \includegraphics[width=.93\columnwidth]{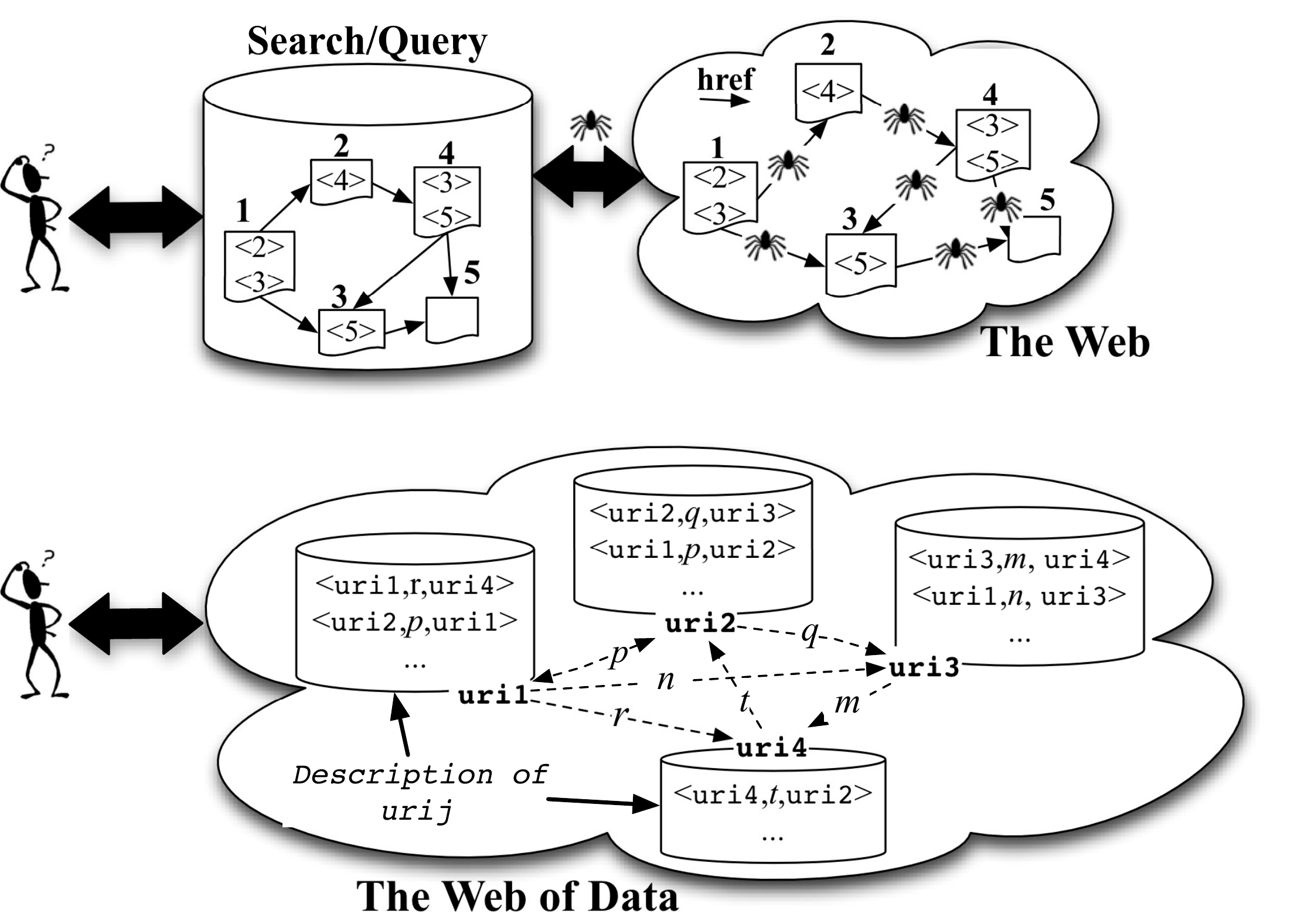}
   \vspace{-.3cm}
    \caption{Classical Web versus Web of Data.
 Size, distributive character, and semantic description 
 of data gives navigation a prominent role.}
  \vspace{-.5cm}
    \label{fig:webdata}
\end{figure}

In this setting, navigation along the nodes of the Web of Data, using the semantics stored in each data source, becomes significant. To model these issues, rather than as a graph, the Web of Data is better represented as a set of nodes plus data describing their semantic structure ``hanging'' from each node (see Fig. \ref{fig:webdata}). This model permits to better express the distributed creation and maintenance of data, and the fact that its structure is provided by dynamical and
distributed data sources. In particular, it reflects the fact that at each moment of time, and for each particular agent, the whole network of data on the Web is unknown~\cite{Mendelzon97}.

This new scenario calls for new models and languages to query and
explore this semantic data space. In particular we highlight three 
functionalities: (1) a new type of navigation emerges as an important feature, in order to traverse sites and data sources; (2) closely tied to it, navigation charts or specifications, that is, semantic descriptions of fragments of the Web; (3) specification of actions one would like to perform over this data ({\em e.g.}, retrieving data, sending messages, etc.) also becomes relevant. Navigation, specifications of regions, and actions appear as part of the basic functionalities for exploring and doing data management over the Web of Data. Ideally, one would like to have a simple declarative language that integrates all of them.
%%
%%
%%%%%
%%
\begin{figure*}[t]
\centering
\vspace{-.4cm}
\includegraphics[width=.85\textwidth]{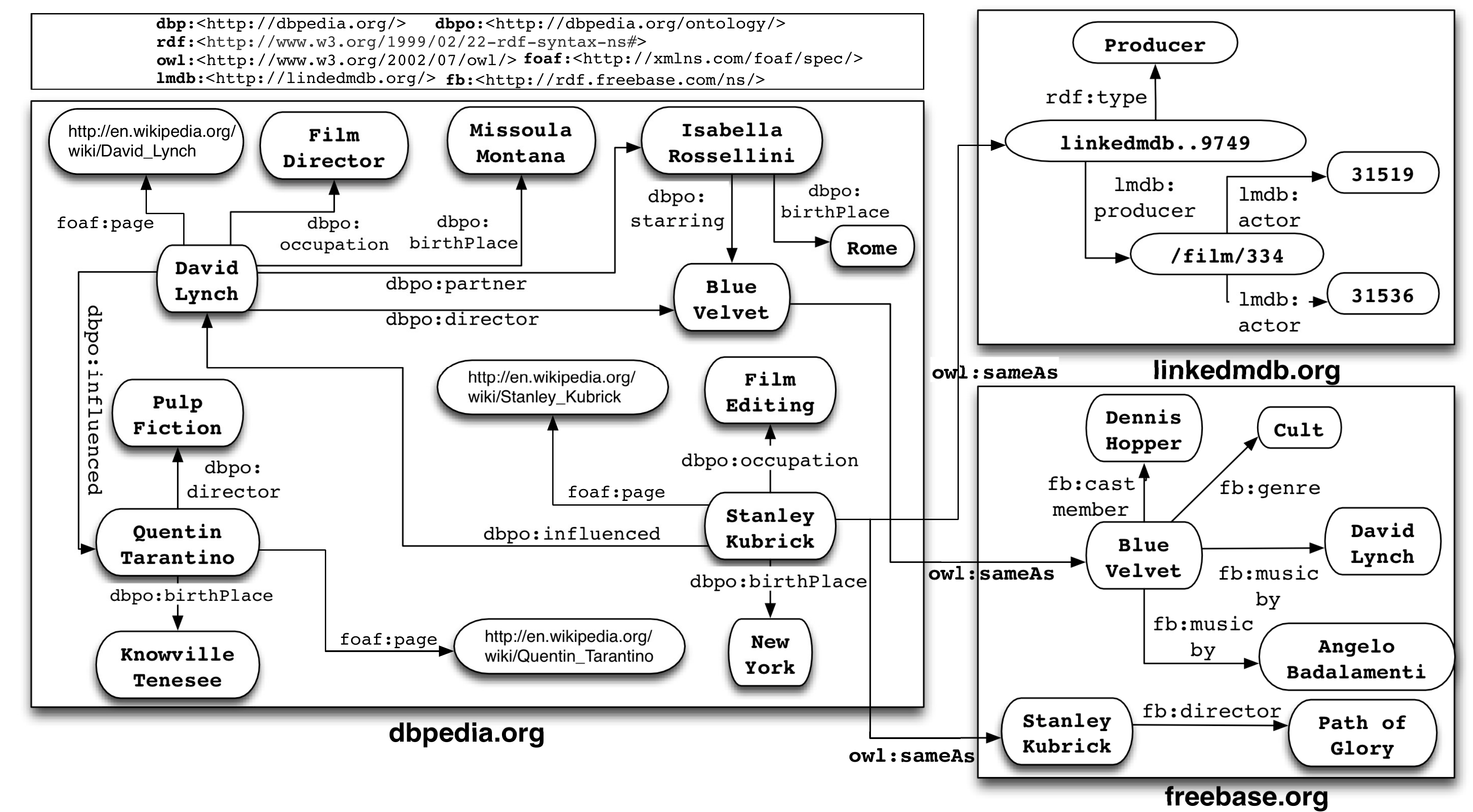}
\vspace{-.5cm}
\caption{An excerpt of data that can be navigated from \texttt{dbpedia:StanleyKubrick}.}
\label{fig:interpretation}
\end{figure*}
%%%
%%
 
 In this paper we present such a language, which we call \nautilod, and show that it can be readily implemented on the current Linked Open Data (LOD) network~\cite{Heath2011}. In fact, we introduce the \swget\ tool that exploits current Web protocols and work on LOD data. Finally, we explore its distributed version and implement an application as proof-of-concept to show its feasibility, potentialities and challenges. 

\medskip
\noindent
\textbf{\nautilod\ by example}. To help the reader to get a more concrete idea of the language, we present some examples using an excerpt of real-world data shown in Fig. \ref{fig:interpretation}. (The formal syntax and semantics is introduced in Section 3).
%%
%%%%%%%%%%%%%%%%%%%%%%%%%%%%%%
%%
\begin{exmp}(\textbf{Aliases via \texttt{owl:sameAs}})
Specify what is predicated from Stanley Kubrick in DBPedia and also consider his possible aliases in other data sources.
\label{example:aliases-relatedness}
\end{exmp}
The idea is to have \texttt{<owl:sameAs>}-paths, which start from Kubrick's URI in DBPedia. Recursively, for each URI \ur\ reached in this way, check in its data source the triples 
$\tuple{\ur, \texttt{owl:sameAs}, \texttt{v}}$. Select all \texttt{v}'s found. Finally, for each of such \texttt{v}, return all URIs \texttt{w} in triples of the form $\tuple{\texttt{v}, \texttt{p}, \texttt{w}}$ found in \texttt{v}'s data source. The specification in \nautilod\ is as follows:
%%%
%%%
\begin{center}
\small
\boxed{
\begin{gathered}
(\texttt{<owl:sameAs>})* / \texttt{<\_>}
\end{gathered}}
\end{center}
where \texttt{<\_>} denotes a wild card for RDF predicates. In Fig. \ref{fig:interpretation},
when evaluating this expression starting from the URI \texttt{dbp:StanleyKubrick}
we get all the different representations of Stanley Kubrick provided
by \texttt{dbpedia.org}, \texttt{freebase.org} and
\texttt{linkedmdb.org}. From these nodes, the expression \texttt{<\_>}
matches any predicate. The final result is: {\scriptsize\{\texttt{dbp:DavidLynch, dbp:New York,} \texttt{dbp:FilmEditing,lmdb:Producer,lmdb:/film/334,fb:Path of Glory,http://en.wikipedia.org/wiki/Stanley\_Kubrick}\}}. Note that the naive search for Kubrick's information in DBPedia, would only give
 {\scriptsize{\texttt{\{http://en.wikipedia.org/wiki/Stanley\_Kubrick, New York, David Lynch, Film Editing\}}}}.

\medskip
\noindent A more complex example, which extends standard navigational languages with actions and SPARQL queries is:
\begin{exmp}
URIs of movies (and their aliases), whose director is more than 50 years old, and has been influenced, either directly or indirectly, by Stanley Kubrick. Send by email the Wiki pages of such directors as you get them.
\label{example:info-films}
\end{exmp}
This specification involves \texttt{influence}-paths and
aliases as in the previous example;
tests over the dataset associated
to a given URI (if somebody influenced by Kubrick is found,
check if it has the right age), a test expressed in \nautilod\ using ASK-SPARQL queries;
and actions to be taken using data form the data source. The \nautilod\ specification is:
\begin{center} \small \boxed{
\begin{gathered}
\texttt{(<dbpo:influenced>)+[Test]/{Act}/<dbpo:director>/} 
\\ \texttt{/(<owl:sameAs>)?}
\end{gathered}
}
\end{center}
where the test and the action are as follows: \\ \texttt{Test=} {\small \texttt{ASK~?p~<dbpo:birthDate>~?y.~FILTER(?y<1961-01-01).}} \\ \texttt{Act=} {\small \texttt{sendEmail(?p)[SELECT ?p WHERE \{?x <foaf:page> ?p.\}]}}.

In the expression, the symbol \texttt{+} denotes
that one or more levels of influence are acceptable, e.g.,
we get directors like David Lynch and Quentin Tarantino. 
From this set of  resources, 
the constraint on the age enforced by the ASK
query is evaluated on the data source associated to 
each of the  resources already matched.
 This filter leaves in this case only
\texttt{dbp:DavidLynch}.
At this point, 
over the elements of this set (one element in this case),
 the action will send via email the page
(obtained from the SELECT query). The action
\texttt{sendEmail}, implemented by an ad-hoc programming procedure,
does not influence the navigation process. 
Thus, the evaluation
will continue from the URI $\ur=$\texttt{dbp:DavidLynch}, 
by navigating the property \texttt{dbpo:director} (found in the dataset $\DD$ 
obtained by dereferencing $\ur$). For example, in $\DD$ we found
the triple $\tuple{\ur,\texttt{dbpo:director}, \texttt{dbp:BlueVelvet}}$. 
Then, from \texttt{dbp:BlueVelvet} we launch the final part of 
the expression, already seen in Example \ref{example:aliases-relatedness}. 
It can be checked that the final result of the evaluation is:
(1)  the set \texttt{\{dbp:BlueVelvet, fb:BlueVelvet}\}, 
that is, data about the movie Blue Velvet
from \texttt{dbpedia.org} and \texttt{freebase.org};
(2) the set of actions done; in this case one email sent.

\medskip
\noindent
\textbf{Contributions of the paper.} The following are main the contributions of this paper:

(1) First:  {\em we define a general declarative specification language},
called \nautilod, whose navigational features exploit regular
expressions on RDF predicates, enhanced with existential tests (based
on ASK-SPARQL queries) and actions. It allows both: to specify a set of
sites that match the semantic description, and to orient the
navigation using the information that these sites provide. Its basic
navigational features are inspired both by \texttt{wget} and
\texttt{XPath}, enhanced with semantic specifications,
using SPARQL to filter paths, and with actions to be performed while navigating. We present a simple syntax, a formal semantics and a basic cost analysis.  %

(2) Second: {\em we implement a version of the language}, 
by developing the application \swget\
that evaluates \nautilod\ expressions in a  centralized form 
(at the distinguished initial node).
Being based on \nautilod, \swget\ permits to perform semantically-driven navigation of the Web of Data as well as retrieval actions. This tool relies on the computational resources of the initial node issuing the command and exploits the Web protocol HTTP. It is readily available on the current Linked Open Data (LOD) network. Its limitation is, of course, the scalability: the traffic of data involved could be high, making the navigation costly. 
%This points to the need of a distributed tool with similar features.
 
(3) Third: {\em we implement} \swget\ {\em in a
distributed environment}. Based on simple assumptions on third parties
(a small application that each server should run to join it, and
that in many ways extends the idea of current {\em endpoints}), we show
the feasibility of such an application that simulates a travelling
agent, and hint at the powerful uses it can have. From this proof-of-concept,
we explore the potentialities of this idea and its challenges.

\medskip
The paper is organized as follows.
Section \ref{sec:preliminaries} provides a quick overview of the Web of Data.
In Section \ref{sec:wod-language} the \nautilod\ language is introduced:
syntax, semantics and its evaluation costs.
In Section \ref{sec:swget} \texttt{swget}, a centralized implementation of \nautilod\ is introduced: its architecture, pseudo-code and experimental evaluation. Section \ref{sec:distributed-swget} deals with the distributed version of \texttt{swget}, showing the feasibility and potentialities
of this application. Section \ref{sec:related-work} discusses related work. Finally, in Section \ref{sec:conclusion} we draw conclusions and delineate future work.
%%%
%%
%%%
%%
\section{Preliminaries: The Web of Data}
\label{sec:preliminaries}
This section provides some background on RDF and Linked Open Data
(LOD) that are at the basis of the Web of Data. For further details the reader can refer to~\cite{bizer09a, Gutierrez2011}.

\smallskip
\noindent
{\bf RDF.}
The Resource Description Framework (RDF) is a metadata model
introduced by the W3C for representing information about resources in
the Semantic Web. RDF is built upon the notion of {\em statement}.
A statement defines the {\em property} $p$ holding between two  resources,
the \textit{subject} $s$ and the \textit{object} $o$. It is denote by $\tuple{s,p,o}$, and thus called  {\em triple} in RDF.
A collection of RDF triples is referred to as an \textit{RDF graph}.
 RDF exploits Uniform Resource Identifiers (URIs) to identify
resources. URIs represent global identifiers in the Web and enable to
access the \textit{descriptions} of resources according to specific protocols
(e.g., HTTP).
%%
%%%%%%%%%%%%%%%%%%%
\subsection{Web of Data - the LOD initiative}
\label{subsec::lod}
The LOD initiative leverages RDF to publish and interlinking resources on the Web. This enables a new (semantic) space called {\em Web of Data}. Objects in this space are \textit{linked} and looked-up by exploiting (Semantic) Web languages and technologies. LOD is based on some principles, which can be seen more as best practices than formal constraints:

(1) Real world objects or abstract concepts must be assigned names on the form of URIs.

(2)  In particular, HTTP URIs have to be used so that people can look them up by using existing technologies.

(3)  When someone looks up a URI, associated information has to be provided in a standard form (e.g., RDF).

(4)  Interconnections among URIs have to be provided by including references to other URIs.

An important notion in this context is that of dereferenceable URI. A dereferenceable URI, represents an identifier of a real world entity that can be used to retrieve a representation, by an HTTP GET, of the resource it identifies. The client can negotiate the format (e.g., RDF, N3) in which it prefers to receive the description.
\subsection{Data in the LOD}
\label{sec:data-lod}
Data in the LOD are provided by sites (i.e., servers), which cover a variety of domains. For instance, \dbpo\ or \texttt{freebase.org} provide cross-domain information, \texttt{geonames.org} publishes geographic information, \texttt{pubmed.org} information in the domain of life-science whereas \texttt{acm.org} covers information
about scientific publications. 

Theoretically in each server resides an RDF triple-store (or a repository of RDF data). In order to obtain information about the resource identified by a URI \ur, a client has to perform an HTTP GET \ur\ request. This request is handled by the Linked Data server, which answers with a set triples.  This is usually said to be the  {\em dereferencing} of \ur.

In the Web of Data, resources are not isolated from one another, in spirit with the fourth principle of LOD, but are linked. The interlinking of these resources and thus of the corresponding sites in which they reside forms the so called \textit{Linked Open Data  Cloud} \footnote{\url{http://richard.cyganiak.de/2007/10/lod/}}.
%%
%%
%%%%%%%%%%%%%%%%%%%%%%%%%%%%%%%%%%%%%%%%%%%%%%%%%%%%%%%%%%%%%%%%%%%
\section {A Navigation Language for the Web of Data}
\label{sec:wod-language}
%%
%%%
%%%

As we argued in the Introduction, there are data management challenges emerging in the Web of Data that need to be addressed. Particularly important are: \textit{(i)} the specification of parts of this Web, thus of semantic fragments of it; \textit{(ii)} the possibility to declaratively specify the navigation and exploit the semantics of data placed at each node of the Web; \textit{(iii)} performing actions while navigating. To cope with this needs, this section presents a {\em navigation language} for the Web of Data, inspired by two non-related languages: \texttt{wget}, a language to automatically navigate and retrieve Web pages; and \texttt{XPath}, a language to specify parts of documents in the world of semi-structured data. We call it {\em Navigational language for Linked Open Data}, \nautilod.

\nautilod\ is built upon \textit{navigational expressions}, based on regular expressions, filtered by tests using ASK-SPARQL queries (over the data residing in the nodes that are being navigated), and incorporating actions to be triggered while the navigation proceeds. \nautilod\ allows to: \textit{(i)} semantically specify collections of URIs; \textit{(ii)} perform recursive navigation of the Web of Data, controlled using the semantics of the RDF data \textit{hanging} from the URIs that are visited (that can be obtained by dereferencing these URIs); \textit{(iii)} perform actions on specific URIs, as for instance, selectively retrieve data from them.

Before presenting the language, we present in Section \ref{sec:model} an abstract data model of the Web of Data. Then we present the syntax of \nautilod\ (Section \ref{sec:wod-syntax}), and the formal semantics (Section \ref{sec:wod-semantics}). Finally, we provide a basic cost model for the complexity of evaluating \nautilod\ expressions.
\subsection{Data  model}
\label{sec:model}
We define a minimal abstract model of the Web of Data to highlight the main features required in our discussion.

Let $\UU$ be the set of all URIs and $\LL$ the set of all literals. We distinguish between two types of triples. \textit{RDF links} $\tuple{s,p,o}\in\UU \times \UU \times \UU$ that encode connections among resources in the Web of Data. {\em Literal triples}, $\tuple{s,p,o}\in\UU \times \UU \times \LL$, which are used to state properties or features of the resource identified by the subject \textit{s}.  Note that the object of a triple, in the general case, can be also a blank node. However, here we will not consider them to simplify the presentation of the main ideas (note also that the usage of blank nodes is discouraged~\cite{Heath2011}). Let $\TT$ be the set of all triples in the Web of Data. The following three notions will be fundamental.
%%%
%%
\begin{defn}[Web of Data $\TT$]
Let\quad $\UU$ and $\LL$ be infinite sets. The Web of Data (over $\UU$ and $\LL$) is the set of triples $\tuple{s,p,o}$ in $\UU \times \UU \times (\UU \cup \LL)$. 
We will denote it by $\TT$.
\end{defn}
\begin{defn}[Description Function $\DD$]
\label{def:description}
A function $\DD:\UU\rightarrow P(\TT)$ 
associates to each URI $u \in \UU$ a subset of triples of $\TT$,
denoted by $\DD(u)$, which is the set of triples obtained by dereferencing $u$.
\end{defn}
%%
% At this point, it is possible to introduce the notion of Web of Data (WoD) instance.
%%%
\begin{defn}[Web of Data Instance $\WW$]
\label{def:global-rdf-graph}
A Web of Data instance is a pair $\WW=\tuple{\UU,\DD}$, 
where $\UU$ is the set of all URIs and $\DD$ is a description function.
\end{defn}
Note that not all the URIs in $\UU$ are dereferenciable. If a URI $\ur \in \UU$ is not dereferenciable then $\DD(\ur)=\emptyset$.
\subsection{Syntax} 
\label{sec:wod-syntax}
\nautilod\ provides a mechanism to declaratively: (i) define \textit{navigational expressions}; (ii) allow {\em semantic control} over the navigation via test queries; (iii) {\em retrieve data} by performing actions as side-effects along the navigational path.

The navigational core of the language is based on regular path expressions, pretty much like Web query languages and XPath. The semantic control is done via existential tests using ASK-SPARQL queries. This mechanism allows to redirect the navigation based on the information present at each node of the navigation path. Finally, the language allows to command actions during the navigation according to decisions based on the original specification and the local information found. 
\begin{table}[ht]
\center
\small
\boxed{
\begin{tabular}{rl}
%%%
% \textbf{Path:}
 $\pathw$ ::= & $ \texttt{pred} \mid (\texttt{pred})^{-1}  \mid \texttt{action} \mid \pathw / \pathw  $ \\
               & $\mid (\pathw)?  \mid (\pathw)*\mid (\pathw | \pathw) \mid \pathw[\texttt{test}]  $ \\
\texttt{pred} ::= & $\textbf{<RDF predicate>} \mid \textbf{<\_>}$ \\
\texttt{test} ::=& \textbf{ASK-SPARQL query} \\
\texttt{action} ::= & $\textbf{procedure[Select-SPARQL query]}$
\end{tabular}
}
\caption{Syntax of the \nautilod\ language.
}
\label{table:syntax}
\end{table}

The syntax of the language \nautilod\ is defined according to the grammar reported in Table \ref{table:syntax}. The language is based on {\em Paths Expressions}, that is, concatenation of base-case expressions built over 
{\em predicates}, {\em tests} and {\em actions}. The language accepts concatenations of basic and complex types of expressions. Basic expressions are predicates and actions; complex expressions are disjunctions of expressions; expressions involving a number of repetitions using the features of regular languages; and expressions followed by a test. The building blocks of a \nautilod\ expression are:
\begin{enumerate}
\item {\em Predicates.} The base case. \texttt{pred} can be an RDF predicate or the wildcard
\texttt{<\_>} used to denote {\em any} predicate.
\item {\em Test Expressions.} A \texttt{test} denotes a query expression. Its base case is an ASK-SPARQL query.  
\item {\em Action Expressions.} An \texttt{action} is a procedural specification of a command (e.g., send a notification message, PUT and GET commands on the Web, etc.), which obtains its \textit{parameters} from the data source reached during the navigation. It is a side-effect, that is, it does not influence the subsequent navigation process.
\end{enumerate}

If restricted to (1) and (2), \nautilod\ can be seen as a declarative language to describe portions of the Web of Data, i.e., set of URIs conform to some semantic specification.
\begin{table*}[htb]
\center
\begin{tabular}{rcl}
$E\llbracket \texttt{<p>}\rrbracket(\ur,\WW)$
  &=& $\{(\ur',\bot) \mid  \tuple{\ur,\texttt{<p>},\ur'} \in \DD(\ur) \}$ \\
$E\llbracket (\texttt{<p>})^{-1} \rrbracket(\ur,\WW)$
 &=& $\{(\ur',\bot) \mid  \tuple{\ur',\texttt{<p>},\ur} \in \DD(\ur) \}$ \\
$E\llbracket \texttt{<\_>}\rrbracket(\ur,\WW)$
  &=& $\{(\ur',\bot) \mid \exists \texttt{<p>}, 
  \tuple{\ur,\texttt{<p>},\ur'} \in \DD(\ur) \}$ \\
$E\llbracket \texttt{act} \rrbracket(\ur,\WW)$
  &=& $\{ (\ur, \texttt{act}) \}$ \\
$E\llbracket \pathw_1/\pathw_2\rrbracket(\ur,\WW)$
  &=&  $\{(\ur'',a) \in E\llbracket \pathw_2\rrbracket(\ur',\WW) : \exists b, (\ur',b) \in E\llbracket \pathw_1\rrbracket(\ur,\WW) \}$\\
$E\llbracket (\pathw)?\rrbracket(\ur,\WW)$
   &=& $\{ (u,\bot)\}\cup E\llbracket \pathw\rrbracket(\ur,\WW)$\\
$E\llbracket (\pathw)*\rrbracket(\ur,\WW)$
  &=& $\{(\ur, \bot)\}\cup \bigcup_{1}^{\infty } E\llbracket \pathw_i \rrbracket(\ur,\WW) \mid \pathw_1=\pathw \wedge \pathw_i=\pathw_{i-1}/\pathw$\\
$E\llbracket \pathw_1|\pathw_2\rrbracket(\ur,\WW)$
   &=& $E\llbracket \pathw_1\rrbracket(\ur,\WW)
        \cup E\llbracket \pathw_2\rrbracket(\ur,\WW)$\\
$E\llbracket \pathw[\texttt{test}]\rrbracket(\ur,\WW)$
  &=& $\{ (\ur',a) \in E\llbracket \pathw\rrbracket(\ur,\WW) :  
   \texttt{test}(\ur')=\texttt{\textbf{true}}\}$\\
\\
$U\llbracket \pathw \rrbracket(\ur,\WW)$ 
   &=&   $\{ v : \exists a, (v,a) \in E\llbracket \pathw \rrbracket(\ur,\WW) \}$  \\
$A \llbracket \pathw \rrbracket(\ur,\WW)$ 
   &=&  
 $\{ \texttt{Exec}(a,v) :  
    (v, a) \in E\llbracket \pathw \rrbracket(\ur,\WW) \}$ \\
\\
Sem$\llbracket \pathw \rrbracket(\ur,\WW)$
  &=& $( U\llbracket \pathw \rrbracket(\ur,\WW),
       A\llbracket \pathw \rrbracket(\ur,\WW))$
\end{tabular}
\caption{Semantics of \nautilod.
The semantics of an expression is composed of two sets: 
(1) the set of URIs of $\WW$ satisfying the specification;
(2) the actions produced by the evaluation of the specification. 
\texttt{Exec}$(a,u)$ denotes the execution of action $a$ over $u$.
$\bot$ indicates the empty
action (i.e., no action).
}
\label{table:semantics}
\end{table*}
\subsection{Semantics} 
\label{sec:wod-semantics}
\nautilod\ expressions are evaluated against a Web of Data instance $\WW$ and a URI \ur\ indicating the starting point of the evaluation. The meaning of a \nautilod\ expression is a set of URIs defined by the expression plus a set of actions produced by the evaluation of the expression. The resulting set of URIs are the leaves in the paths according to the \nautilod\ expression, originating from the seed URI \ur.

For instance, the expression \texttt{type}, evaluated over \ur, will return the set of URIs $\ur_k$ reachable from \ur\ by ``navigating'' the predicate  \texttt{type}, that is, by inspecting triples of the form $\tuple{\ur, \texttt{type} ,\ur_k}$ included in $\DD(\ur)$. Similarly, the expression \texttt{type}[\q] will filter, from the results of the evaluation of \texttt{type}, those URIs $\ur_k$ for which the query \q\ evaluated on their descriptions $\DD(\ur_k)$ is true. Finally, the evaluation of an expression \texttt{type}[\q]/\texttt{a} will return the results of \texttt{type}[\q] {\em and} perform the action \texttt{a} (possibly using some data from $\DD(\ur_k)$).

The formal semantics of \nautilod\ is reported in Table \ref{table:semantics}. The fragment of the language without actions follows the lines of  formalization of XPath by Wadler~\cite{Wadler99}. Actions are treated essentially as side-effects and evaluated while navigating. Given and expression, a Web of Data instance $\WW= \tuple{\UU, \DD}$, and a seed URI \ur\, the semantics has the following modules:
\begin{itemize}
\item $E\llbracket \pathw \rrbracket(\ur,\WW)$: Evaluates the set of URIs selected by the navigational expression \pathw\ starting from the URI \ur\ in the Web of Data instance $\WW$. Additionally, it collects the actions associated to each of such URIs.
\item  $U\llbracket \pathw \rrbracket(\ur,\WW)$: Defines the set of
URIs specified by the expression \pathw\ when forgetting the actions.
\item $A\llbracket \pathw \rrbracket(\ur,\WW)$: Executes the actions
specified by the evaluation of the navigational expression \pathw. 
\item $Sem\llbracket \pathw \rrbracket(\ur,\WW)$: 
Outputs the meaning of the expression \pathw, namely,
the ordered pair of two sets: the set of URIs specified by the
evaluation of \pathw; and
the set of actions performed according to this information.
\end{itemize}
{\em Note on some decisions made:} 
 Any sensible real implementation can benefit from giving 
an order to the elements of the output action set. As far as the formal semantics, at this stage we assumed that actions are independent from one another and that the world $\WW$ is static during the evaluation (to avoid to overload our discussion with the relevant issue of
synchronization, that is at this point orthogonal to the current proposal). Thus, we decided to denote the actions produced by the evaluation of an expression as a set. It is not difficult to see that one could have chosen a list as the semantics for output actions. 
\subsection{Evaluation of Costs and Complexity}
\label{sec:complexity-swget}
We present a general analysis of costs and complexity of the evaluation of \nautilod\ expressions over a Web of Data instance $\WW$.
We can separate the costs in three parts, where $E$ are expressions,
$\overline{E}$ action-and-test-free expressions, $A$ actions and $T$ tests:
\begin{equation}
\label{a}
  cost(E,\WW) = cost(\overline{E},\WW) + cost(A)+cost(T).
\end{equation}
Since actions do not affect the navigation process we can treat their cost separately. Besides, in our language, tests are ASK-SPARQL queries having a different structure from the pure navigational path expressions of the language. Even in this case we can treat their cost independently. 

\smallskip
\textbf{Actions.} \nautilod\ is designed for acting on the Web of Data.
In this scenario, the cost of actions has essentially two components: \textit{execution} and \textit{transmission}. The execution cost boils down to the cost of evaluating the SELECT SPARQL query that gives the action's parameters. As for transmission costs, a typical example is the \texttt{wget} command, where the cost is the one given by the GET data command.

\smallskip
\textbf{Action-and-test-free.}  This fragment of \nautilod\ can be considered essentially as the PF fragment of {\em XPath} (location paths without conditions), that is well known to be (with respect to combined complexity)
$\mathbf{NL}$-complete under $\mathbf{L}$-reductions (Thm. 4.3, \cite{Gottlob2003}).
 The idea of the proof is simple: membership in $\mathbf{NL}$ follows from 
the fact that we can guess the path while we verify it in time $\mathbf{L}$.
The hardness essentially follows from a reduction from the directed graph reachability problem. Thus we have:
%%%
%%%
\begin{theorem}  With respect to combined complexity, the action-and-test-free fragment of \nautilod\ is $\mathbf{NL}$-complete under $\mathbf{L}$-reductions. 
\end{theorem}
%%%
%%%
{\em Combined} refers to the fact that 
the input parameters are the expression size and the data size.
Note that what really matters is not the whole Web (the data),
but only the set of nodes reachable by the expression. Thus
it is more precise to speak of expression size plus set-of-visited
nodes size. The worst case is of course the whole size of the Web.

\smallskip
\textbf{Tests.} The evaluation of tests (i.e., ASK-SPARQL queries) has a cost. This cost is well known and one could choose particular fragments of SPARQL to control it \cite{Perez09}. However, tests will possibly reduce the size of the set of nodes visited during the evaluation. Thus the $cost(\overline{E},\WW)$ has to be reduced to take into account the effective subset of nodes reachable thanks to the filtering performed by the tests. Let $\WW_{T}$ be $\WW$ when taking into account this filtering. We have:
\begin{equation}
\label{b}
  cost(E,\WW) = cost(\overline{E},\WW_{T}) + cost(A)+cost(T).
\end{equation}
Section \ref{sec:eval-expr} will discuss some examples on real world data by underlining the contribution of each component of the cost.

\smallskip
\textbf{Final Considerations.} In a distributed setting, with partially unknown information and a network of almost unbound size, the notion ``cost of evaluating an expression e" appears less significant than in a controlled centralized environment. In this scenario, a more pertinent question seems to be: ``given an amount of resources \textit{r} and the expression \textit{e}, how much can I get with \textit{r} satisfying \textit{e} ?''. This calls for optimizing (according to some parameters) the navigation starting from a given URI $u$, according to equation (\ref{b}).
\section{Implementation of \nautilod} 
\label{sec:swget}
This section deals with \swget, a tool implementing \nautilod. The tool \swget\ implements all the navigational features of \nautilod, a set of actions centred on retrieving data, and adds (for practical reasons) a set of ad-hoc options for further controlling the navigation from a network-oriented perspective (e.g., size of data transferred, latency time) that today's are not yet found as RDF statements.

\swget\ has been implemented in Java and is available as: \textit{(i)} a developer release, which includes a command-line tool that is easily embeddable in custom applications; \textit{(ii)} an end user release, which features a GUI. Further details, examples, the complete syntax along with the downloadable versions are available at the \swget's Web site \footnote{\url{http://swget.wordpress.com}}. 
\subsection{Architecture}
\begin{figure}[h]
\centering
\vspace{-.4cm}
\includegraphics[width=.5\textwidth,clip]{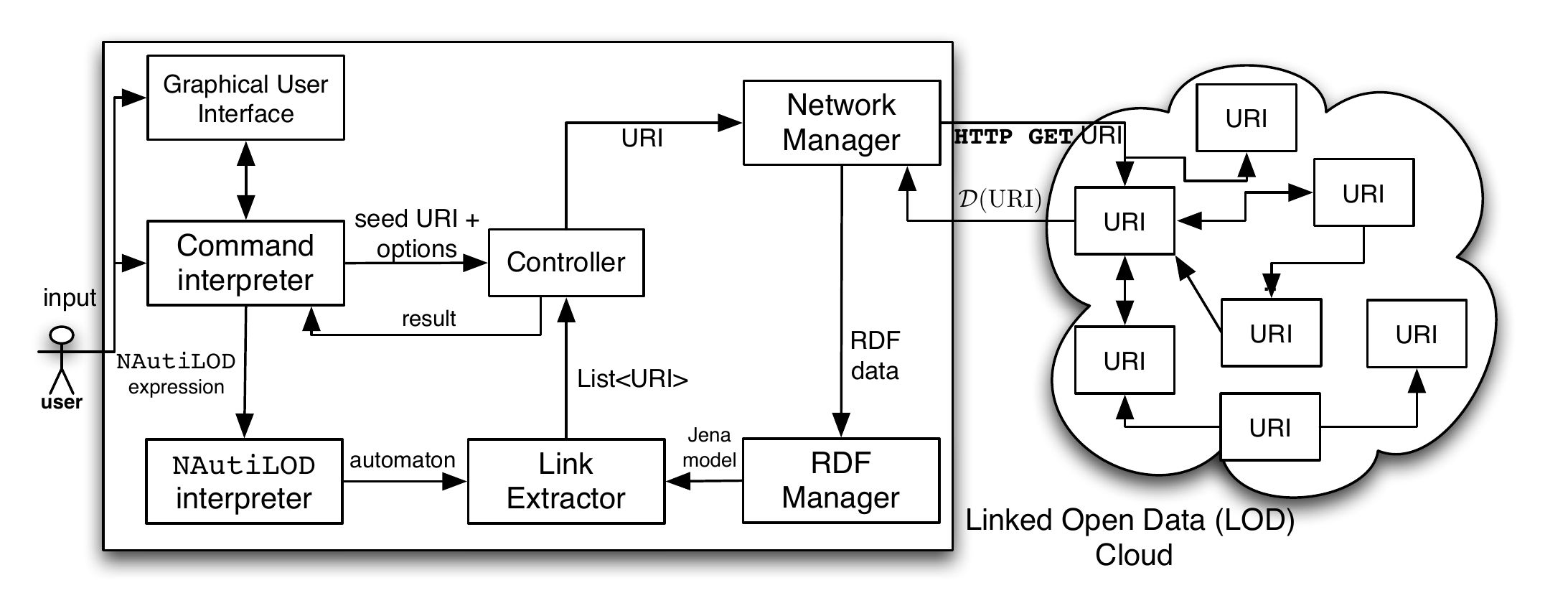}
\vspace{-1cm}
\caption{\swget\ architecture and scenario.}
\label{fig:swget-architecture}
\end{figure}

The high level architecture of \swget\ is reported in the left part of Fig. \ref{fig:swget-architecture}. The \textit{Command interpreter} receives the input, i.e., a seed \texttt{URI}, a \nautilod\ expression and a set of options. The input is then passed to the \textit{Controller} module, which checks if a network request is admissible and possibly passes it to the \textit{Network Manager}. A request is admissible if it complies with what specified by the \nautilod\ expression and with the network-related navigation parameters (see Section \ref{sec:nav-parameters}). The \textit{Network Manager} performs HTTP GET requests to obtain streams of RDF data. These streams are processed for obtaining Jena RDF models, which will be passed to the \textit{Link Extractor}. The \textit{Link Extractor} takes in input an automaton constructed by the \nautilod\ \textit{interpreter} and selects a subset of outgoing links in the current model according to the current state of the automaton. The set is given to the \textit{Controller Module}, which starts over the cycle. The execution will end either when some navigational parameter is satisfied or when there are no more URIs to be dereferenced.
%%%%
%%
%%
%%
\subsubsection{Network-based controlled navigation}
\label{sec:nav-parameters}
\nautilod\ is designed to semantically control the navigation. However, it can be the case that a user wants to control the navigation also in terms of network traffic generated. A typical example is a user running \swget\ from a mobile device with limited Internet capabilities. This is why \swget\ includes features to add more control to the navigation through the parameters reported in Table \ref{table:network-params}. Each option is given in input to \swget\ as a pair $\langle \texttt{param},\texttt{value} \rangle$.
\vspace{-.5cm}
\begin{table}[htb]
\scriptsize
\caption{Network params to control the navigation}
\label{table:network-params}
\center
\begin{tabular}{|c|c|p{3.5cm}|}\hline
\textbf{Parameter}&\textbf{Value}&\textbf{Meaning}\\\hline
\texttt{maxDerTriples}&int& max. number of triples allowed in each dereferencing\\\hline
\texttt{saveGraph}&boolean&Save the graphs dereferenced\\\hline
\texttt{maxSize}&int&traffic limit (in MBs)\\\hline
\texttt{timeoutDer}&long&connection time-out\\\hline
\texttt{timeout}&long&total time-out\\\hline
\texttt{domains}&List<String>&trusted servers\\\hline
\end{tabular}
\end{table}

To illustrate a possible scenario where the navigation can be controlled both from a semantic and network-based perspective consider, the following example.
\begin{exmp}(\textbf{Controlled navigation})
Find information about Rome, starting from its definition in DBPedia and includes other possible definitions of Rome linked to DBPedia but only if their description contains less than 500 triples and belongs to DBPedia, Freebase or The New York Times.
\label{example:controlled-navigation}
\end{exmp}
\begin{center}
\boxed{
\begin{gathered}
\swget\ <\texttt{dbp:Rome> (<owl:sameAs>)*} \ \texttt{-saveGraph}\\
\texttt{-domains \{dbpedia.org,rdf.freebase.com,} \\ \texttt{data.nytimes.com\}}  \ \texttt{-maxDerTriples 500}
\end{gathered}}
\end{center}
The command, besides the \nautilod\ expression, contains the \texttt{-domains} and \texttt{-maxDerTriples} parameters to control the navigation on the basis of the trust toward information providers and the number of triples, respectively.
%%%
%%
%%
\subsection{Evaluating \nautilod\ expressions}
\label{sec:eval-expr}
Given a \nautilod\ expression $\e$ it is possible to build an automaton that can recognize \nautilod\ expressions. The transitions between states of the automaton implements the navigation process.
%%%%
%%%
%%
\subsubsection{The \swget\ Navigation Algorithm}
\label{sec:nav-algo}
The \swget\ controlled navigation algorithm is reported in Algorithm \ref{code:swget}. Moreover, Table 4 describes the high level primitives used in the pseudo-code to interact with the automaton.
\vspace{-.2cm}
\begin{algorithm}[ht]
\scriptsize
\SetAlgoLined
  \SetKwInOut{Input}{Input}\SetKwInOut{Output}{Output}\SetKwFunction{checkNetParams}{checkNet}\SetKwFunction{navigate}{navigate}
%\begin{algorithmic}[1]
\Input{\texttt{e}=\nautilod\ expression; \texttt{seed}=URI; \texttt{par}=\texttt{Parms<n,v>}}
  \Output{set of URIs and literals conform to \texttt{e} and \texttt{par};}
  \BlankLine
  \algsetup{linenosize=\scriptsize,linenodelimiter=, }
 \begin{algorithmic}[1]

\STATE $\a=\texttt{buildAutomaton(e)}$;
\STATE \texttt{addLookUpPair(\texttt{seed}, \texttt{a.getInitial()})};
\WHILE{($\exists$ \texttt{p=<uri,state>} to look up and \texttt{checkNet(par)=OK})}
    \STATE\texttt{desc=getDescription(p.uri)};
    \IF {(\texttt{a.isFinal(p.state}))} \STATE \texttt{addToResult(p.uri)};
    \ENDIF
	
    \IF {(not \texttt{alreadyLookedUp(p)})}
\STATE	\texttt{setAlreadyLookedUp(p)};
    \IF {(\texttt{t=getTest(p.state)}$\neq\emptyset$ and \texttt{evalT(t,desc)=true})}
\STATE    \texttt{s=a.nextState(p.state,t))};
\STATE   		\texttt{addLookUpPair(p.uri,s)};
\ENDIF
        \IF {(\texttt{act=getAction(p.state)$\neq\emptyset$})}
\STATE    \textbf{if}(\texttt{evalA(act.test,desc})) \textbf{then} \texttt{exeC(act.cmd)};
\STATE    \texttt{s=a.nextState(p.state,act))};
\STATE   		\texttt{addLookUpPair(p.uri,s)};
\ENDIF
\STATE    \texttt{out}=\textbf{navigate}(\texttt{p,a,desc});
\FOR {(each URI pair \texttt{p'=<uri,state>} in \texttt{out})}
\STATE		\texttt{addLookUpPair(p')};
\ENDFOR
	\FOR{(each literal pair \texttt{lit=<literal,state>} in \texttt{out})}
  \IF{(\texttt{a.isFinal(lit.state)})}
\STATE \texttt{addToResult(lit.literal)};   
\ENDIF
\ENDFOR
\ENDIF
\ENDWHILE
\STATE return Result\;
\end{algorithmic}
\caption{\swget\ pseudo-code}
\label{code:swget}
\end{algorithm}
\vspace{-1cm}
\begin{function}[ht]
\scriptsize
  \SetKwInOut{Input}{Input}\SetKwInOut{Output}{Output}\SetKwFunction{alreadyChecked}{alreadyChecked}
  \Output{List of \texttt{<uri,state>} and \texttt{<literal,state>}} 
  \BlankLine
  \algsetup{linenosize=\scriptsize,linenodelimiter=, }
 \begin{algorithmic}[1]
\FOR {(each \texttt{pred} in \texttt{a.nextP(p.state)})}
\STATE	\texttt{nextS=a.nextState(p.state,pred)}\;
\STATE	\texttt{query= \textquotedbl SELECT ?x WHERE \\\{\{ ?x pred p.uri\} UNION\{ p.uri pred ?x\}\}\textquotedbl}\;
	\FOR {(each \texttt{res} in \texttt{evalQ(query, \texttt{desc})})}
			\STATE \texttt{addOutput(res,nextS)}\;
	\ENDFOR
\ENDFOR
\STATE return Output\;
\end{algorithmic}
\caption{navigate(\texttt{exp},\a,\texttt{desc})}
\end{function}
\vspace{-.5cm}
%%%
\begin{table}[htb]
\label{table:primitives}
\scriptsize
%%\center
\caption{Primitives for accessing the automaton.}
\begin{tabular}{|p{1.7cm}|p{5.8cm}|}\hline
\textbf{Primitive}&\textbf{Behaviour}\\\hline\hline
\texttt{getInitial()}&returns the initial state $q_0$\\\hline
\texttt{nextP(q)}&returns the set $\{\sigma \mid \delta(q,\sigma)=q_1\}$ of tokens (i.e., predicates) enabling a transition from $q$ to $q_1$\\\hline
\texttt{getTest(q)}&returns the test to perform into the current automaton state\\\hline
\texttt{getAction(q)}&returns the action to perform into the current automaton state\\\hline
\texttt{nextState(q,$\sigma$)}&returns the state that can be reached from \texttt{q} by the token $\sigma$\\\hline
\texttt{isFinal(q)}&returns \texttt{TRUE} if \texttt{q} is an accepting state\\\hline
\end{tabular}
\end{table}

The algorithm takes as input a \textit{seed} \texttt{URI}, a \nautilod\ expression and a set of network parameters, and returns a set of URIs and literals conform to the expression and the network parameters. For each URI involved in the evaluation, possible tests (line 9) and actions (line 12) are considered. 

The procedure \textbf{navigate} is exploited to extract links (line 3) from a resource identified by \texttt{p.uri} toward other resources. According to the Linked Data initial proposal~\cite{Berners2006} [section on browsable graphs] \texttt{p.uri} may appear either as the \textit{subject} or the \textit{object} of each triple.
\begin{figure*}[htb]
\scriptsize
\subfigure[Time (secs)]{
%%\center
\hspace{-.41cm}
%%1
\includegraphics[scale=.48]{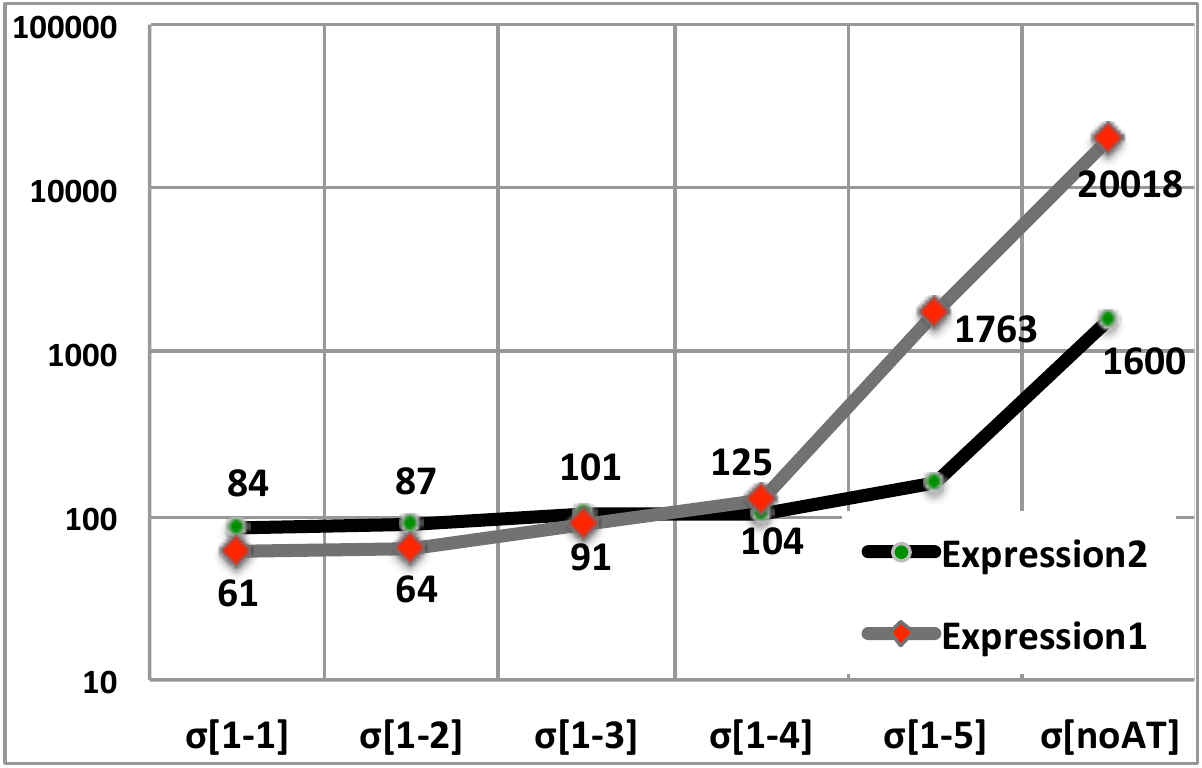}
}
\hspace{-.31cm}
\subfigure[\#Dereferenced URIs]{
\includegraphics[scale=.48]{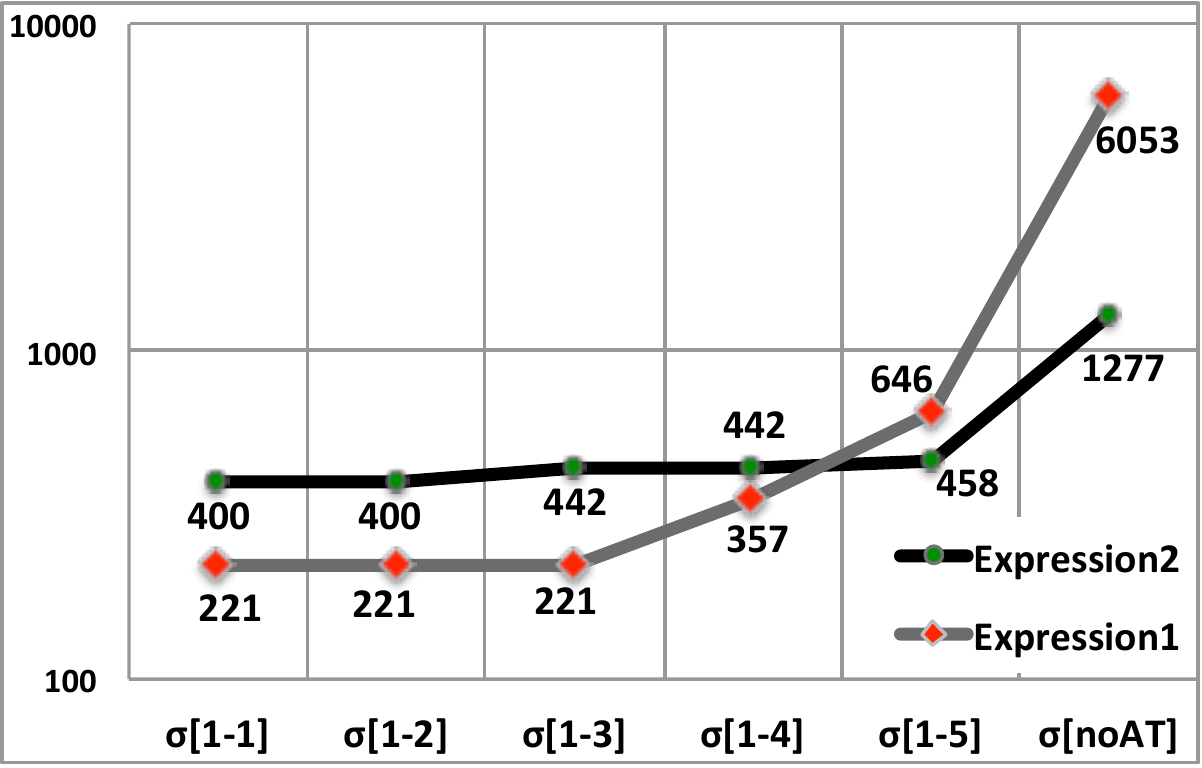}
}
\hspace{-.31cm}
\subfigure[\# Triples retrieved]{
\includegraphics[scale=.48]{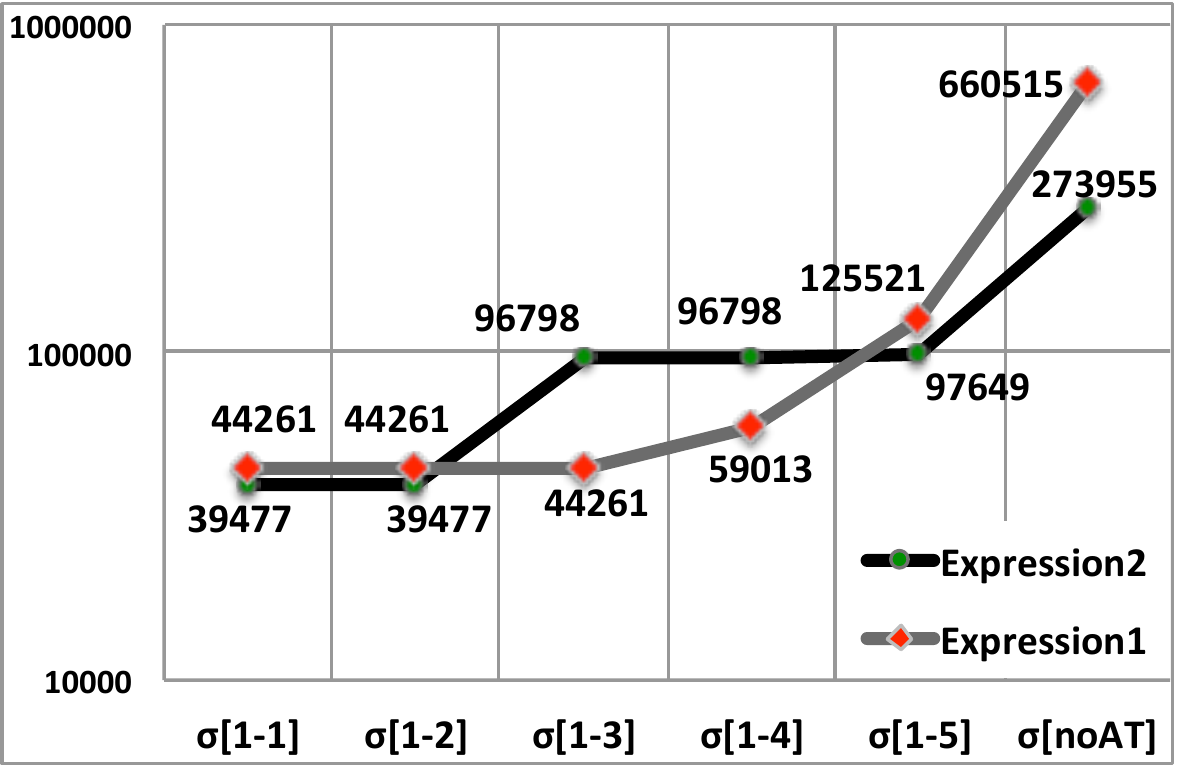}
}
\caption{Evaluation of \swget. Each expression has been executed 4 times. Average results are reported.}
\label{fig:evaluation}
\end{figure*}
\subsection{Experimental Evaluation}
\begin{figure}[h]
\centering
%\vspace{-.4cm}
\includegraphics[width=\columnwidth,clip]{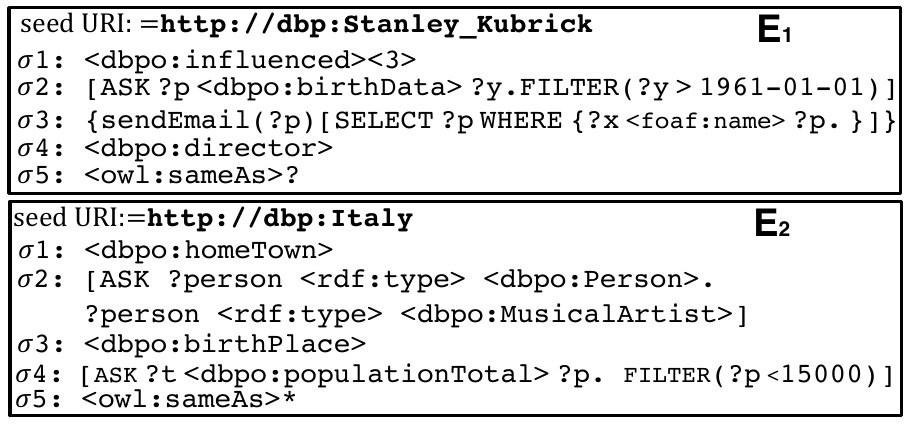}
\vspace{-.8cm}
\caption{Expressions used in the evaluation}
\label{fig:dataset}
\end{figure}

\setcounter{subfigure}{0}
%%%%
\label{sec:evaluation}
To show real costs of evaluating the different components of \swget\ expressions over real-world data, we choose two complex expressions (shown in Fig. \ref{fig:dataset}) to be evaluated over the Linked Open Data network. We report the results of \swget\ in terms of execution time ($t$), URIs dereferenced ($d$) and number of triples retrieved ($n$). Each expression has been divided in 5 parts (i.e., $\sigma_i, i\in \{1..5\}$). They have been executed as whole (i.e., $\sigma_{[1-5]}$) and as action-and-test-free expressions (i.e., $\sigma_{[noAT]}$), which correspond to $\overline{E}_1$ and $\overline{E}_2$, respectively (see Section \ref{sec:complexity-swget}). Moreover, the various sub-expressions (i.e., $\sigma_{[1-i]}, i \in \{1..4\}$) have also been executed. This leads to a total of 12 expressions. For each expression, the corresponding \textit{sub-Web} has been locally retrieved. That is, for each reachable URI the corresponding RDF graph has been locally stored. The aim of the evaluation is to investigate how the various components in the cost model presented in Section \ref{sec:complexity-swget} affect the parameters $t$, $d$ and $n$.
%%
% %%
%%
%%%%
%%
%%%%

The results of the evaluation, in logarithmic scale, are reported in Fig. \ref{fig:evaluation}(a)-(c). In particular, in the X axis are reported from left to right: the 4 sub-expressions, the full expression (i.e., $\sigma_{[1-5]}$)  and the action-and-test-free expression  (i.e., $\sigma_{[noAT]}$). Note that in some cases, the number of results is higher than the number of dereferenced URIs reported because not all the results were dereferenceable URIs.

The first expression ($E_1$) starts by finding people influenced by Stanley Kubrick up to a level 3 (subexpr. $\sigma_{[1-1]}$). This operation requires about 61 secs., for a total of 221 URIs dereferenced. On the description of each of these 221 URIs, an ASK query is performed to select only those entities that were born after the 1961 (subexpr. $\sigma_{[1-2]}$). The execution time of the queries is of about 4 secs. (i.e., $\simeq$ 0.02 secs., per query). Hence, 31 entities have been selected. At this point, an action is performed on the descriptions of these 31 entities by selecting their \texttt{<foaf:name>} to be sent via email (subexpr. $\sigma_{[1-3]}$). In total, the select, the rendering of the results in an HTML format and the transmission of the emails cost about 25 secs. The navigation continues from the 31 entities before the action to get movies through the property \texttt{<dbpo:director>} (subexpr. $\sigma_{[1-4]}$). The cost is of about 34 secs., for a total of 136 movies. Finally, for each movie only one level of possible additional descriptions is searched by the \texttt{<owl:sameAs>} property (the whole expr. $\sigma_{[1-5]}$) whose cost is 1638 secs., for a total of 409 new URIs available from multiple servers (e.g., \texttt{linkedmdb.org}, \texttt{freebase.org}) of which only 289 were dereferenceable.
%%%%

By referring to the cost model in Section \ref{sec:complexity-swget} we have that $cost(E_1,\WW)=cost(\overline{E}_1,\WW_{T_1})+cost(A_1)+cost(T_1)=1763$. Here, the factor $cost(A_1)\simeq25$ secs., whereas $cost(T_1)\simeq4$ secs., and $cost(\overline{E}_1,\WW_{T_1})\simeq1738$ secs. If we consider the test-and-action-free expression executed over the whole Web of Data (i.e., $\WW$), we have that $cost(\overline{E}_1,\WW)\simeq 20018$ secs. Note that the ASK queries costs about 4 secs., and permits to reduce the portion of the Web of Data navigated by $\overline{E}_1$, which enables to save about $20018-1738=18280$ secs. Such a larger difference in the execution times is justified by the fact that the 222 initial URIs, selected by $\sigma_{[1-1]}$ are not filtered in the case of $(\overline{E}_1,\WW)$ and then cause an larger amount of paths to be followed at the second level. Indeed, the total number of dereferenced URIs for $(\overline{E}_1,\WW)$ is 6053 while for $(\overline{E}_1,\WW_{T_1})$ is 646 with about 660K triples retrieved in the first case and 125K in the second case. 

The second expression ($E_2$) starts by navigating the property \texttt{<dbpo:homeTown>} to find entities living in Italy (subexpr. $\sigma_{[1-1]}$) with an execution time of about 84 secs., and a total of 400 dereferenced URIs, one seed and 399 URIs of entities. On the description of each of these 399 URIs, an ASK query filters entities that are of type \texttt{<dbpo:Person>} and \texttt{<dbpo:MusicalArtist>} (subexpr. $\sigma_{[1-2]}$). Hence, 399 ASK queries are performed for a total of about 3.8 secs., with an average time per query of 0.01 secs., to select 156 entities. For these entities, the navigation continues through the property \texttt{<dbpo:birthPlace>} to find the places where these people were born (subexpr. $\sigma_{[1-3]}$), which costs about 101-87=14 secs. In total, 43 new URIs have been reached. The navigation continues with a second ASK query to select only those places in which live less than 15000 habitants (subexpr. $\sigma_{[1-4]}$). The cost of performing 43 ASK queries on the results of the previous step is of about 3 secs. Here 5 places are selected. Finally, for each of the 5 places additional descriptions are searched by navigating the \texttt{<owl:sameAs>} property (the whole expr. $\sigma_{[1-5]}$). This allows to reach a total of 29 URIs, some of which are external to \dbpo. The cost for this operation is of about 57 secs. 

As for the cost, we have $cost(E_2,\WW)=cost(\overline{E}_2,\WW_{T_2})+cost(A_2)+cost(T_2)=161$. The factor $cost(A_2)=0$ since $E_2$ does not contain any action whereas $cost(T_2)\simeq6$ secs. Hence, $cost(\overline{E}_2,\WW_{T_2})=155$ secs. The cost of the test-and-action-free expression (i.e., $\overline{E}_2$) over $\WW$ is $cost(\overline{E}_2,\WW)\simeq 1600$ secs., for a total of 1277 dereferenced URIs. This is because the expression is not selective since it performs a sort of "semantic" crawling only based on RDF predicates. In fact, the number of triples retrieved (see Fig. \ref{fig:evaluation}(c)) is almost three times higher than in the case of the expression with tests. By including the tests, the evaluation of $\overline{E}_2$ is $1445$ secs., faster.
\section{A Proposal for Distributed \swget}
\label{sec:distributed-swget}
This section presents and overview of Distributed \swget\ (\dswget), which has the peculiarity that the processing \nautilod\ expressions occurs in a cooperative manner among LOD information providers. The tool has been implemented and tested on a local area network.
%%%
%%%
\subsection{\dswget: making LOD servers cooperate }
\label{sec:d-swget}
\swget\ enables controlled navigation but it heavily relies on the client that initiates the request. However, one may think of the Linked Data servers storing RDF triples as to peers in a Peer-to-Peer (P2P) network, where links are given by URIs in RDF triples. For instance $\tuple{\texttt{dbp:Rome}, \texttt{owl:sameAs}, \texttt{fb:Rome}}$ links \texttt{dbpedia.org} with \texttt{freebase.org}. Indeed, there are some differences w.r.t. a traditional P2P network. First, Linked Data servers are less volatile than peers. Second, it is reasonable to assume that the computational power of Linked Data servers is higher than that of a traditional peers. This enables to handle a higher number of connections with the associated data.

Our proposal is to leverage the computational power of servers in the network to cooperatively evaluate \swget\ commands. This enables to drastically reduce the amount of data transferred. In fact, data is not transferred from servers to the client that initiates the request (in response to HTTP GETs). Servers will exchange \swget\ commands plus some metadata and operate on their data locally. This can be achieved by installing on each server in the network a \dswget\ engine and coordinating the cooperation by an ad-hoc distributed algorithm. 
%%%
%%
\vspace{-.5cm}
\begin{table}[h]
\scriptsize
\caption{Primitives of \dswget}
%%\center
\begin{tabular}{|p{1.7cm}|p{5.8cm}|}\hline
%%\label{table:primitives}
\textbf{Primitive}&\textbf{Behaviour}\\\hline\hline
\texttt{sendResults}&sends to the original client (partial) results, which are URIs (line 14) and literals (line 18)\\\hline
\texttt{fwdToServers}&forwards to other servers, the initial client address, the \nautilod\ expression and a set of pairs \texttt{<URI, A State>}. For each pair, the computation on a URI will be started from the corresponding \texttt{A state}\\\hline
\end{tabular}
\label{tab:prim-dswget}
\end{table}
%%
%%%

 A \dswget\ command is issued by a \dswget\ client to the server to which the \textit{seed} URI belongs. Each server involved in the computation will receive, handle and forward commands and results by using the Procedure \textbf{\ref{proc:handle}}. Note that in this procedure there are calls to some primitives reported in Table 4 and to the function \textbf{navigate} described in Section \ref{sec:nav-algo}. The specific primitives needed by \dswget\ are reported in Table \ref{tab:prim-dswget}.
\begin{procedure}[ht]
\scriptsize
 \SetKwInOut{Input}{Input} 
%\begin{algorithmic}[1]
\Input{\texttt{client\_id}=address of the client; \texttt{e}=\nautilod\ expression; \texttt{URIs}=set of pairs \texttt{<\texttt{URI,A state}>}; \texttt{metadata}=additional data (e.g., current state of the automaton, request id)}
%%%
 \BlankLine
   \algsetup{linenosize=\scriptsize,linenodelimiter=, }
\begin{algorithmic}[1]
\STATE \texttt{\a=\texttt{buildAutomaton(\e})};
\FOR{(each \texttt{p=<uri,state>} in \texttt{URIs})}
   \STATE \texttt{desc=getDescription(p.uri)}; //local call no deref. needed\
     \IF {(not \texttt{alreadyLookedUp(p)})}
\STATE	\texttt{setAlreadyLookedUp(p)};
    \IF {(\texttt{t=getTest(p.state)}$\neq\emptyset$ and \texttt{evalT(t,desc)=true})}
\STATE    \texttt{s=a.nextState(p.state,t))};
\STATE   		\texttt{addLookUpPair(p.uri,s)};
\ENDIF
             \IF {(\texttt{act=getAction(p.state)$\neq\emptyset$})}
\STATE    \textbf{if}(\texttt{evalA(act.test,desc})) \textbf{then} \texttt{exeC(act.cmd)};
\ENDIF
\STATE    \texttt{out}=\textbf{navigate}(\texttt{p,a,desc});
\FOR {(\textbf{each} URI pair \texttt{p'=<uri,state>} in \texttt{out})}
\IF{(\texttt{a.isFinal(p'.state)})}
\STATE        \texttt{addtoResults(p'.uri)}; 
\STATE       \textbf{else} \texttt{addLookUpPair(p')};
\ENDIF
\ENDFOR
   \FOR{(\textbf{each} literal pair \texttt{lit=<literal,state>} in \texttt{out})}
 \IF{(\texttt{a.isFinal(lit.state)})}
\STATE \texttt{addToResult(lit.literal)};
\ENDIF
\ENDFOR   
\ENDIF 
\ENDFOR
\STATE \texttt{sendResults(client\_id)}\;
\STATE \texttt{fwdToServers}(\texttt{client\_id},\texttt{e})\;
\end{algorithmic}
\caption{handle(\texttt{client\_id,e,URIs, metadata})}
\label{proc:handle}
%%%%%%%
%%%%%%%
\end{procedure}
\begin{figure*}[ht]
\centering
\includegraphics[height=.38\linewidth,clip]{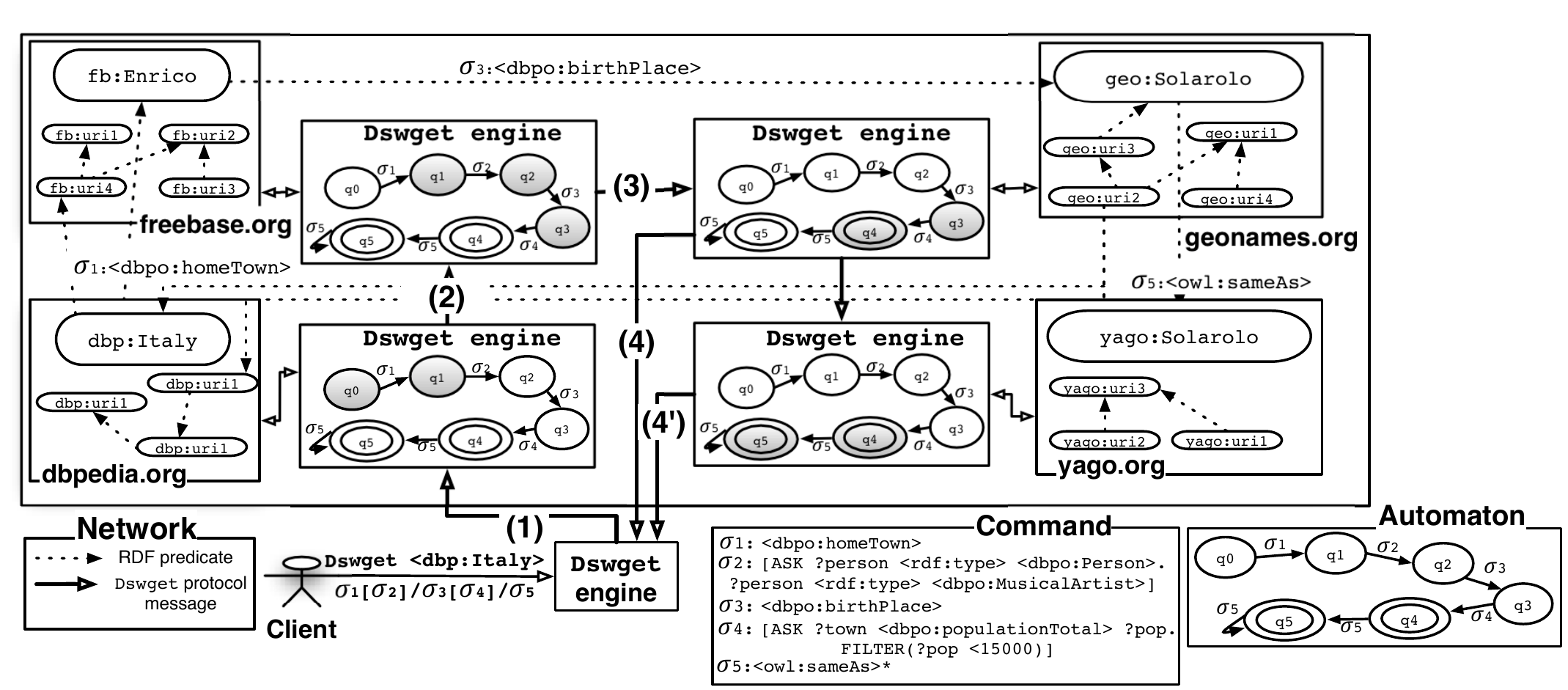}
\vspace{-.5cm}
\caption{Distributed \dswget\ interaction scenario.}
\label{fig:dist-swget-scenario}
\end{figure*}
%%%
%%
\subsubsection{A Running example}
\label{sec:example-dswget}
To see an example of how \dswget\ works, consider the following request originated from a \dswget\ client:
%%
%%%%%%%
\vspace{-.1cm}
%%%%%%%
\begin{exmp}(\textbf{\dswget}) Starting from DBPedia, find cities with less than 15000 persons, along with their aliases, in which musicians, currently living in Italy, were born.
\label{example:distributed-swget}
\end{exmp}
%%%%%%%
\vspace{-.1cm}
%%%%%%%
%%
The \dswget\ command is reported in Fig. \ref{fig:dist-swget-scenario}, which also reports a possible \dswget\ interaction scenario. On each linked data server a \dswget\ engine has been installed. 
Each server exposes a set of dereferenceable URIs for which the corresponding RDF descriptions are available. RDF data enables both internal references (e.g., \texttt{dbp:Rome} and \texttt{dbp:uri1}) and external ones (e.g., \texttt{fb:Enrico} and \texttt{geo:Paris}). 

In Fig. \ref{fig:dist-swget-scenario} references between URIs are represented by dotted arrows. When not explicitly mentioned, it is assumed that the reference occurs on a generic predicate. The automaton associated to this expression, having  \texttt{q4} and \texttt{q5} as accepting states, is also reported. The state(s) of the automaton on which a server is operating is(are) reported in grey. \dswget\ protocol messages have been numbered to emphasize the order in which they are exchanged.

The command along with some medatada (e.g., the address of the client) is issued by the client's \dswget\ engine toward the server to which the seed URI belongs (i.e., \texttt{dbpedia.org} in this example). The \dswget\ engine at this server, after locally building the automaton, starts the processing of the \nautilod\ expression at the state \texttt{q0}. It obtains from its local RDF store, the description of Rome $\DD(\texttt{dbp:Rome})$ and looks for URIs having \texttt{dbpo:hometown} as a predicate. In Fig \ref{fig:dist-swget-scenario}, the URI \texttt{fb:Enrico} satisfies this pattern. The \dswget\ engine at \texttt{dbpedia.org} performs the first transition of state, that is, $\delta(\texttt{q0},\sigma1)=\texttt{q1}$. The automaton does not reach a final state, and then the process has to continue. Since the URI \texttt{fb:Enrico} belongs to another server, the \dswget\ engine at \dbpo, communicates with that at \fbo\ by seeding the initial \nautilod\ expression, the URI for which \fbo\ is involved in the computation (i.e., \texttt{fb:Enrico}) and the current state of the automaton. In the case in which multiple URIs have to be sent, they are packed together in a unique message.

With a similar reasoning the request reaches the \dswget\ engine at \geoo, which checks if it is possible to reach the next state of the automaton starting from the URI passed by \fbo. It has to check on $\DD(\texttt{geo:Solarolo})$ if the query represented by $\sigma_4$ can be satisfied, that is, whether this city has less than 15K habitants. Then, the state \texttt{q4} is reached, which is a final state. The \dswget\ engine at \geoo\ contacts directly the \dswget\ engine of the client that issued the request and send the result (i.e., the URI \texttt{geo:Solarolo}). The address of the client is passed at each communication among \dswget\ engines.

Note that the automaton has another final state, that is, \texttt{q5} that can be reached if there exist some triples in $\DD(\texttt{geo:Solarolo})$ having an \texttt{owl:sameAs} predicate. Such a triple is $\tuple {\texttt{geo:Solarolo, owl:sameAs, yago:Solarolo}}$. Therefore, the \dswget\ engine at \geoo\ sends to the engine at \yagoo\ the URI in the object of this triple, the expression and the current state of the automaton. Here, as the automaton is in a final state, the \dswget\ engine sends to the client the result and continues the process. In this case since in $\DD(\texttt{yago:Solarolo})$ there are no more triples having \texttt{owl:sameAs} as predicate, the process ends.
 %%
%%%%%%%
\vspace{-.2cm}
%%%%%%%
\subsection{\dswget\ Design issues: an overview}
\label{sec:design-issues}
%%%%
In designing \dswget\ several issues, typical of the distributed systems, have been faced. Here we briefly report on the main of them without getting into too technical details.

In the Web of Data, a client in order to get information about a resource issues an HTTP GET request toward the HTTP server where the resource is hosted. In the standard case, the HTTP protocol offers a blocking semantics for its primitives, which means that once a request is issued the client has to wait for an answer or until a time-out. In \dswget, since engines exchange messages and data in a P2P fashion, a blocking semantics for communications would block the whole execution. To face this issue, specific asynchronous communication primitives and a \textit{job delegation} mechanism have been implemented. With job delegation we mean that the sending \dswget\ engine delegates part of the execution and evaluation of a (sub)\nautilod\ expression to the receiving engine(s). In this respect, since a request, through the mechanism of \textit{job delegation} is spread among multiple \dswget\ engines it is necessary to handle the termination of requests to avoid to keep consuming resources in an uncontrolled way. \dswget\ tackles this issue from two different perspectives:

(1) \textit{Loop detection}: each \dswget\ engine keeps track, for each request, of each URI along with the state of the automaton on which it has been processed.

(2) \textit{Termination}: this problem can be addressed by each \dswget\ engine which, for each request it receives informs the client that initially issued the request about the fact that it has operated on this request and whether it has delegated other \dswget\ engines. Then, the client can keep track of the list of the active engines on a particular request.  The \dswget\ engine may additionally send back to the client the state of the automaton on which it is operating, thus enabling the client to know how far the execution is from a final state.
%%
%%%%%%%
\vspace{-.2cm}
%%%%%%%
%%
\section{Related Work}
\label{sec:related-work}
%%%%
Many of the ideas underlying our proposal have been around in particular settings. We owe inspiration to several of them. 

 {\em Navigation and specification} languages of nodes in a graph have a deep
research background. Nevertheless, most of its developments assume that
data is stored in a central repository (e.g.
Web query languages~\cite{Florescu1998}, {\em XPath},
%% %%
%%%
navigational versions of SPARQL~\cite{Perez2010,Alkhateeb09}).
They were inspiration for the navigational core of \nautilod.

   Specification (and retrieval) of collections of sites was early
addressed, and a good example is the well known tool \texttt{wget}. Besides
 being non-declarative, it is restricted to almost purely
syntactic features. At semantic level,
Hart et al. ~\cite{Isele2010} proposed LDSpider, a crawler for
the Web of Data able to retrieve RDF data by following RDF links
according to different crawling strategies.
They have little flexibility and are not declarative.
 The execution philosophy of
\texttt{wget} was  a source of inspiration for the incorporation
of actions into \nautilod\ and to the design of \texttt{swget}.

 Distributed data management has been explored and implemented
by P2P and similar approaches~\cite{Valduriez04}. For RDF,  RDFPeers~\cite{Cai04} and YARS2 uses P2P to answer RDF
queries. Systems for distributed query processing on the Web have also been
devised, e.g. DIASPORA ~\cite{Ramanath00}. 
Our distributed version of \swget\ borrows some ideas from these
approaches.

Finally, it is important to stress the fact that there is a solid body of
work on query processing and navigation on the Web of Data. 
Three lines of research can be identified:

\textit{(1)} Load the
\textit{desired} data into a single RDF store (by \textit{crawling}
the LOD or some sub-portions) and process queries in a
centralized way. There is a large list of Triple Stores~\cite{Hose11}.
%% %%
There have been also developments in indexing techniques for semantic data.
 Swoogle ~\cite{Ding2004}, Sindice ~\cite{Oren2008} and Watson
~\cite{dAquin2011} among the most successful. 
Recently, Hart et al. ~\cite{Harth10} proposed an approximate 
index structure for summarizing the content of Linked Data sources.

\textit{(2)} Process the queries in a distributed form by using a federated query processor. DARQ~\cite{quilitz2008} and FedX~\cite{Schwarte2011} provide mechanisms for transparently query answering on multiple query services. The
query is split into sub-queries that are forwarded to the individual
data sources and their result processed together. An evaluation of federated query approaches can be found in  ~\cite{Haase10}.

\textit{(3)} Extend SPARQL with navigational features. The SERVICE feature of SPARQL 1.1 and proposals like the one of Hartig et al.~\cite{Hartig09} extend the scope of SPARQL queries with navigational features~\cite{Hartig09,Hartig11}. 
The system SQUIN, based on {\em link-traversal}, 
a query execution paradigm that discover on the fly data sources relevant for the query, permits to automatically navigate to other sources while executing a query.

As it can be seen, our approach has a different departure point:
it focuses on {\em navigational} functionalities, thus departing from
querying as in \textit{(2)}; emphasizes  specification of autonomous distributed sources, as opposed to \textit{(1)}; uses SPARQL querying to enhance navigation, 
while \textit{(3)} proceeds in the reverse direction; and incorporates actions that in some sense generalize procedures implicit in the evaluation over the Web
({\em e.g.}, ``get data'' in crawlers and  ``return data'' in query languages).
%%%%%%%
%%%%%%%
%%%
\section{Conclusions and Future Work}
\label{sec:conclusion}
We presented a language to navigate, specify fragments and perform actions on the Web of Data. It explicitly exploits the semantics of the data ``stored'' at each URI. We implemented it in a centralized setting to run over real-world data, namely the LOD network, showing the benefits it can bring. We also developed a distributed version as proof-of-concept of
its feasibility, potentialities and challenges.
  
The most important conclusion we can draw from this research
and development is that the semantics given by RDF specifications
{\em can be used} with profit to navigate, specify places and
actions on the Web of Data. We presented a language that can be
used as the basis for the development of agents that get data; navigate
and report while navigating; and that can work immediately
over LOD.  

A second relevant finding we would like to report here,
are the limitations found to take full advantage of the language 
and tools we developed. They refer essentially to (1) lack of standards
in the sites regarding the dereferencing of data; (2) lack of standard 
RDF metadata regarding properties of the sites themselves 
(e.g., provenance, summary of contents, etc.);
(3) weak infrastructure to host delegation of execution and 
evaluation (of the language) to permit distribution. Tackling these
issues can be considered as our wish list to leverage the Web of Data.
%%%
\vspace{-.3cm}
{\footnotesize
\bibliographystyle{plain}
\bibliography{biblio}}
%%%
\end{document}